\documentclass[twocolumn,showpacs,preprintnumbers,amsmath,amssymb]{revtex4}

\usepackage{graphicx}
\usepackage{dcolumn}
\usepackage{bm}

\begin{document}

\preprint{APS/123-QED}

\title{ Scalar potential model of the CMB radiation temperature}

\author{ John C. Hodge }
\altaffiliation[Also at ]{XZD Corp., 3 Fairway St., Brevard, NC, 28712, scjh@citcom.net}
\email{jch9496@blueridge.edu}
\affiliation{Blue Ridge Community College, 100 College Dr., Flat Rock, NC 28731-7756}

\date{\today}

\begin{abstract}
A derivation of a theoretical, time average, cosmic microwave background (CMB), Planckian temperature $V$ of the universe remains a challenge.  A scalar potential model (SPM) that resulted from considerations of galaxy cells is applied to deriving a value for $V$.  The heat equation is solved for a cell with the boundary conditions of SPM Source and Sink characteristics, with simplified cell characteristics, and with zero initial temperature.  The universe is a collection of cells.  The CMB radiation is black body radiation with the cells acting as radiators and absorbers.  Conventional thermodynamics is applied to calculate $V = 2.718$\ldots K.  The temperature and matter content of cells are finely controlled by a feedback mechanism.  Because time is required for matter to flow from Sources to Sinks, the radiation temperature of cells cycles about $V$ after an initial growth phase.  If the universe is like an ideal gas in free expansion and is not in thermal equilibrium, then the pressure and volume follow the measured CMB temperature $v_\mathrm{m} = 2.725 \pm 0.002$ K.  Therefore, increasing $v_\mathrm{m} >V$ equates to an expansion pressure on matter and expanding volume. 
\end{abstract}

\maketitle

\section{ INTRODUCTION}

A derivation of a theoretical, time average, cosmic microwave background (CMB), Planckian temperature $V$ of the universe and an explanation of the deviation of the measured temperature $v_\mathrm{m} = 2.725 \pm 0.002$ K  \citep{benn,math2} from $V$ remains a challenge.\footnote{The choice of symbols conforms to \citet{cars}}  The COBE result showed the CMB radiation spectrum is very close to a pure Planck function.

The standard cosmological, ``big bang'' cosmological models (BBM) view the CMB as a relic of an era of the universe when energy density was very high.  As the universe expanded, the temperature declined adiabaticly.  Therefore, the BBM of the CMB temperature is inversely proportional to the cosmological scale factor such that $v_\mathrm{m}$ is constant in all directions and redshifts.  

In inflationary models, a scalar field causes the early, accelerated expansion of space.  The inflationary universe scenario solves the horizon and flatness problems and the related entropy problem.  Inflation models are also models for the origin of the large--scale structure of the universe and predicted an almost scale-invariant spectrum of cosmological fluctuations.  Inflation models have serious conceptual problems \{see \citet[and references therein]{bran} for a summary\}.  \citet{riess} found evidence for a transition at redshift $z$ = 0.46$\pm$0.13 from a cosmic deceleration to a cosmic acceleration and for a ``cosmic jerk'' at the transition.

The ``Steady State'' models (SSM) posit the universe has always existed and is continually expanding and contracting.  Therefore, energy and matter must be continually being created in intergalactic space and the temperature of the universe must be continually changing \citep{narl}.  The uniformity of the cosmic microwave background (CMB) temperature is considered to have falsified SSM.

A Cyclic Universe model (CU) has recently been reintroduced \citep{stei}.  The CU posits the universe continually expands and contracts with bounces in a four dimensional scalar field.  A major component of CU is that events in previous cycles help shape events in following cycles.

The scalar potential model (SPM) suggests the existence of a massless scalar potential $\rho$ field \citep[and references therein]{hodg}.  Matter and $\rho$ originates from Sources and goes to Sinks.  The SPM was created to be consistent with the morphology--radius and the intragalactic medium cluster observations of galaxies.  Several differences among galaxy types suggest that Sources are located in spiral galaxies and that Sinks are located in early type, lenticular, and irregular galaxies.  The SPM suggests Source and Sink galaxy pairs, triples, groups, and clusters (herein ``cells'') are organized with Sources surrounding the Sinks \citep{aaro2,cecc,huds,lilj,rejk}.  Because the distance between galaxies is larger than the diameter of a galaxy, the Sources were considered as point (monopole) Sources.  The gradient of the $\rho$ field exerts a force $F_\mathrm{s}$ on matter that is repulsive of matter.  In Source galaxies, $F_\mathrm{s}$ repels matter from the galaxy. In Sink galaxies, $F_\mathrm{s}$ repels matter into the galaxy.  The B--band luminosity $L_\mathrm{\epsilon}$ of matter in Source galaxies was considered proportional to the Source strength $\epsilon$.  Therefore, the matter and $\rho$ emitted by a Source are proportional to $\epsilon$.  For the sample galaxies, the ratio of $L_\mathrm{\epsilon}$ to the B--band luminosity $L_\mathrm{\eta}$ of Sink galaxies approaches $2.7 \pm 0.1$.  The SPM was applied to redshift and discrete redshift measurements \citep{hodg}.  In volumes close to a Source, $\rho \propto D^{-1}$, where $D$ is the distance to a Source.  

In this Paper $\rho \propto D^{-1}$ and the intragalactic medium cluster observations suggests the diffusion (heat) equation applies to the flow of energy and matter from Sources to Sinks.  Because the matter ejected from Source galaxies to Sink galaxies is related to $\epsilon$, the feedback control mechanism of matter in a cell and, therefore, of cell radiation temperature $v_\mathrm{l}$ must be in the Sink.  The heat equation is solved with the boundary conditions of SPM Source and Sink characteristics, with simplified cell characteristics, and of zero initial temperature.  The universe is a collection of cells.  The CMB radiation is black body radiation with the cells acting as radiators and absorbers.  Conventional thermodynamics is applied to calculate $V= 2.718$\ldots K.  The $v_\mathrm{l}$ and matter content of cells are finely controlled by a feedback mechanism.  Because time is required for matter to flow from Sources to Sinks, the $v_\mathrm{l}$ cycles about $V$ after an initial growth phase.

The object of this article is to examine the CMB radiation temperature within the context of the SPM.  In section~\ref{sec:model}, the SPM $v_\mathrm{l}$ calculation equation is developed and $V$ is calculated.  The discussion and conclusion is in Section~\ref{sec:disc}.

\section{\label{sec:model}Model}

Posit energy $Q_\mathrm{in}$ is injected into our universe (U) through Source portals from hot, thermodynamic reservoirs (HRs).  The macro thermodynamic processes were considered the same for all Sources.  The $Q_\mathrm{in}$ flows away from the Source.  Some matter remains near the Source as a galaxy and some matter is removed from the Source galaxy by $F_\mathrm{s}$.  Gravitational forces $F_\mathrm{g}$ cause the matter removed from galaxies to concentrate in a Source--less galaxy.  Eventually, enough matter becomes concentrated to initiate a Sink.  Because a minimum amount of matter around a Sink is required to initiate a Sink, a minimum amount of energy $Q_\mathrm{k}$ and matter in a cell is also required.  The Sink ejects energy $Q_\mathrm{out}$ out of U through Sink portals to cold, thermodynamic reservoirs (CRs).  The Sink strength $\eta$ depends on the amount of matter in the local Sink volume.  This is a negative feedback mechanism that tightly controls the energy $Q_\mathrm{u}=Q_\mathrm{in} -Q_\mathrm{out}$ in U.  Therefore, 
\begin{equation}
Q_\mathrm{u} \propto \int _0 ^\mathrm{now} \left( \sum_{i=1}^{N_\mathrm{sources} } \vert \epsilon_i \vert -\sum_{k=1}^{N_\mathrm{sink} } \vert \eta_k \vert \right) \mathrm{d}t
\label{eq:1},
\end{equation}
where $t$ is time since the start of U; $i$ and $k$ are indexes; $N_\mathrm{sources}$ and $N_\mathrm{sink}$ are the total number of Sources and Sinks, respectively, in U; and $\vert \, \vert$ means ``absolute value of''.

Thermodynamic equilibrium for U is when 
\begin{equation}
\sum_{i=1}^{N_\mathrm{sources} } \vert \epsilon_i \vert=\sum_{k=1}^{N_\mathrm{sink} } \vert \eta_k \vert
\label{eq:2}.
\end{equation}
Therefore, if $Q_\mathrm{u}$ is larger than the thermodynamic equilibrium value, $|\eta|$ increases which reduces $Q_\mathrm{u}$.  Conversely, if $Q_\mathrm{u}$ is smaller than the thermodynamic equilibrium value, $|\eta|$ decreases which increases $Q_\mathrm{u}$.

The observation from earth of the multiple Source and Sink galaxies requires the HR to be common for all Sources.  Otherwise, conceptual difficulties similar to the domain-wall problem, the horizon problem, and the flatness problem of BBM occur.  Therefore, the energy from all Sources is causally correlated and coherent.

Because there is a distance between Source and Sink, the matter requires time to move from the Source to the Sink, time to cool, time to penetrate the elliptical galaxy to the Sink, and time to change $\eta$.  Therefore, the cells are not in internal, thermal equilibrium.  

For simplicity, consider only one cell and posit: 
(1) The cell consists of a distribution of Sources around the core of Sinks.  The core Sinks were considered a monopole (point) Sink.  
(2) A Gaussian surface may be constructed enclosing the volume wherein all matter flows to the Sink core.  The Gaussian surface is the border of the cell.\footnote{This is a redefinition of a cell.  \citet{hodg} defined a cell with equal Source and Sink strengths.}  
(3) The temperature $v$ within a cell is a function of distance $x$ from the Sink core and $t$ [$v = v(x,t)$].
(4) The volume of the cell may be characterized by a linear dimension $l$ wherein $l$ is the $x$ of the Gaussian surface and a minimum.  The cells are not required to share boundaries.  The transparency of intercell space supports this assumption.  
(5) The matter in a cell is gained and lost only at the Sources and Sinks of the cell, respectively.  That is, the matter flux across the Gaussian surface is zero at all points of the surface.  
(6) Only radiation and $\rho$ may leave the cell.  
(7) The $x$ to the outermost Source is less than $l$.  
(8) The $v$ is proportional to the matter density in a cell.  
(9) The intragalactic medium cluster observations and $\rho \propto D^{-1}$ suggests the diffusion (heat) equation applies to the flow of matter from Sources to Sinks.  
(10) The initial temperature of U is zero everywhere.  
(11) The Sink feedback control mechanism is the amount of matter around the Sink that controls the amount of radiation $Q(t)$ per unit area per unit time emitted from the cell through the Gaussian surface.  Because the matter transport from the Sources to the Sink core is considerably slower than the speed of light, the matter transport and cooling (conductivity $K$) are the time determining factors of $Q(t)$.   
(12) Because only $v_\mathrm{l}= v(l,t)$ was of concern, the initial condition of the distribution of the Sources and Sinks was ignored.  Therefore, the $v(x,t)$ for values of $x \neq l$ was not calculated.  
(13) The $Q(t)$ is proportional the departure of $v_\mathrm{l}$ from $V$.  Thus, $Q(t)=C(V-v_\mathrm{l})$, where $C$ is a function of the rate of matter input of the Sources in a cell and was considered a constant.  
(14) The radiant energy and $\rho$ from other cells influences $K$.  Because only one cell was considered, $K$ was considered a constant.  
(15) The boundary conditions are 
\begin{eqnarray}
-K \frac{\mathrm{d}{v}(0,t)}{\mathrm{d}x} & = & C (V-v_\mathrm{l}), \nonumber \\
v(x,0) &= & 0
\label{eq:4}.
\end{eqnarray}

The solution of the heat equation for $v_\mathrm{l}$ with these boundary conditions has been performed \{see \citet[\S 15.8, pp. 407-412]{cars}\}.

\begin{figure}
\includegraphics[width=0.5\textwidth]{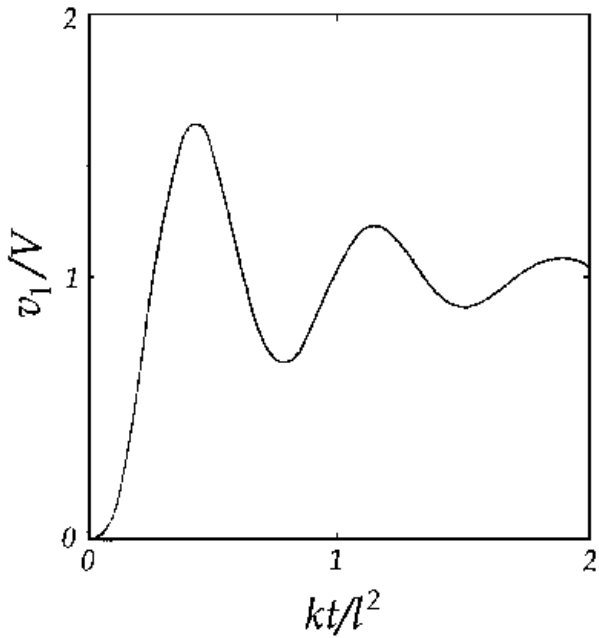}
\caption{\label{fig:1} Behavior of $v_\mathrm{l}$ with feedback control for intermediate values of $kl$. }
\end{figure}

Figure~\ref{fig:1} is a plot of $v_\mathrm{l} /V$ versus $k t/ l^2$ for  a stable value of $k l$, where $k=C/K$ is a positive constant.  

U is the sum of all cells plus the $\rho$ and radiation in the space among cells.  There is no other external energy in U.  

Posit each cell is a radiator like in a black box and all the cells are at the same $v_\mathrm{l}$.  The redshift of photons is caused by a loss of energy from the photon to the universe caused by the $\rho$ field \citep{hodg}.  The lost photon energy must remain in U and be reabsorbed by U.  Full thermalization requires such an emission and absorption process.  Therefore, one of two possibilities exists: (1) All cells were formed at the same time and follow identical evolution paths.  That is, there is a universal time clock.  (2) A communication exists to equalize the temperature of each cell with other cells by means of a feedback mechanism.  For example, the $\rho$ field from a neighboring cell may change the $F_\mathrm{s}$ in a cell that changes the rate of matter transport, hence $k$.  The latter may provide a mechanism for the ``cosmic jerk'' suggested by \citet{riess}. 

When $v_\mathrm{l}>V$, there is excess $Q_\mathrm{u}$.  As $v_\mathrm{l}$ decreases to values less than $V$, the excess $Q_\mathrm{u}$ is removed from U.  Therefore, $v_\mathrm{l} $ converges (``hunts'') to $V$ after a number of cycles that depend on $k l$.  If the value of $k l$ is too low, $v_\mathrm{l} < V$ always.  If $k l$ is too high, the hunting will diverge and the universe will be unstable.  The findings of \citet{riess} suggest $v_\mathrm{l}$ is oscillating about $V$ after $Q_\mathrm{k}$ is established.  Therefore, the process of increasing $Q_\mathrm{u}$ above $Q_\mathrm{k}$ is reversible and U is behaving as a thermodynamic, Carnot engine at 100\% efficiency.  

The Kelvin temperature scale is defined for a Carnot engine such that the ratio of two Kelvin temperatures are to each other as the energies (heats) absorbed and rejected by U, 
\begin{equation}
\frac{Q_\mathrm{hr}}{Q_\mathrm{u}} = \frac{Q_\mathrm{u}}{Q_\mathrm{cr}} =\frac{V}{V_\mathrm{cr}} = \frac{V_\mathrm{hr}}{V}
\label{eq:5},
\end{equation}
where $Q_\mathrm{hr}$ and $Q_\mathrm{cr}$ are the energy in CR and HR, respectively; $V_\mathrm{cr}$ and $V_\mathrm{cr}$ are the time average temperature of the CR and HR, respectively; and the units of $V$, $V_\mathrm{cr}$, and $V_\mathrm{hr}$ are Kelvin.

The amount of heat in each reservoir is proportional to the amount of heat in the previous reservoir.  Also, if the zero point of the Kelvin scale for U is defined as $V_\mathrm{cr}$, then 
\begin{equation}
V = \mathrm{e\,K} =2.718\ldots \, \mathrm{K}
\label{eq:8}.
\end{equation}

\section{\label{sec:disc}Discussion and conclusion}

Within the context of the SPM, the puzzling cosmic acceleration \cite{ riess98,perl99,astier05} and cosmic deceleration \cite{riess} may be explained if the cells are considered as an ideal gas in free expansion.  Because the cells are not in thermal equilibrium ($v_\mathrm{l} \neq V$), both the $\rho$ pressure on matter and volume of the universe follow $v_\mathrm{l}$.  Therefore, increasing $v_\mathrm{l} >V$ equates to an expansion pressure on matter and expanding volume. 

A scalar potential model (SPM) that derived from considerations of cells is applied to the theoretical, time average, radiation temperature $V$ of the universe.  The heat equation is solved with the boundary conditions of SPM Source and Sink characteristics, with simplified cell characteristics, and of zero initial temperature.  The universe is a collection of cells.  The CMB radiation is black body radiation with the cells acting as radiators and absorbers.  Conventional thermodynamics is applied to calculate $V = 2.718$\ldots K.  The temperature and matter content of cells are finely controlled by a feedback mechanism.  Because time is required for matter to flow from Sources to Sinks, the temperature of cells cycles about $V$ after an initial growth phase. 

\begin{acknowledgments}

I acknowledge and appreciate the financial support of Maynard Clark, Apollo Beach, Florida, while I was working on this project.
\end{acknowledgments}



\end{document}